\documentclass[letterpaper]{article}
\usepackage{aaai25}  
\usepackage{times}  
\usepackage{helvet}  
\usepackage{courier}  
\usepackage[hyphens]{url} 
\usepackage{graphicx}
\usepackage{natbib} 
\usepackage{caption}
\usepackage{amsmath}
\usepackage{amsfonts}
\usepackage{booktabs}   
\usepackage{siunitx}    
\usepackage{caption}   
\usepackage{algorithm}
\usepackage{algorithmic}
\nocopyright

\usepackage{newfloat}
\usepackage{listings}
\DeclareCaptionStyle{ruled}{labelfont=normalfont,labelsep=colon,strut=off} 
\floatstyle{ruled}
\newfloat{listing}{tb}{lst}{}
\floatname{listing}{Listing}
\title{When the Past Misleads: Rethinking Training Data Expansion Under Temporal Distribution Shifts}
\author {
    Chengyuan Yao\textsuperscript{\rm 1},
    Yunxuan Tang\textsuperscript{\rm 2},
    Christopher Brooks\textsuperscript{\rm 2},
    Rene F. Kizilcec\textsuperscript{\rm 3},
    Renzhe Yu\textsuperscript{\rm 1}
}
\affiliations {
    \textsuperscript{\rm 1}Columbia University\\
    \textsuperscript{\rm 2}University of Michigan\\
    \textsuperscript{\rm 3}Cornell University\\
    cy2706@tc.columbia.edu, yunxuant@umich.edu, brooksch@umich.edu, kizilcec@cornell.edu, renzheyu@tc.columbia.edu
}

\begin{document}

\maketitle

\begin{abstract}
Predictive models are typically trained on historical data to predict future outcomes. While it is commonly assumed that training on more historical data would improve model performance and robustness, data distribution shifts over time may undermine these benefits. This study examines how expanding historical data training windows under covariate shifts (changes in feature distributions) and concept shifts (changes in feature-outcome relationships) affects the performance and algorithmic fairness of predictive models.
First, we perform a simulation study to explore scenarios with varying degrees of covariate and concept shifts in training data. Absent distribution shifts, we observe performance gains from longer training windows though they reach a plateau quickly; in the presence of concept shift, performance may actually decline. Covariate shifts alone do not significantly affect model performance, but may complicate the impact of concept shifts. In terms of fairness, models produce more biased predictions when the magnitude of concept shifts differs across sociodemographic groups; for intersectional groups, these effects are more complex and not simply additive.
Second, we conduct an empirical case study of student retention prediction, a common machine learning application in education, using 12 years of student records from 23 minority-serving community colleges in the United States. We find concept shifts to be a key contributor to performance degradation when expanding the training window. Moreover, model fairness is compromised when marginalized populations have distinct data distribution shift patterns from their peers.
Overall, our findings caution against conventional wisdom that "more data is better" and underscore the importance of using historical data judiciously, especially when it may be subject to data distribution shifts, to improve model performance and fairness.
\end{abstract}

\begin{links}
    \link{Code}{https://github.com/AEQUITAS-Lab/Distribution-Shift-AIES-2025}
\end{links}

\section{Introduction}
Machine learning applications have been widely deployed to facilitate decision making in social sectors such as healthcare and education~\cite{dixon2024unveiling,BROBY2022145,sghir2023recent}. In developing machine learning models, there is a common assumption that larger training datasets would lead to better performance and greater model generalizability. The logic behind this assumption is intuitive: more data covers more diverse underlying patterns, which can help reduce both bias and variance in predictive estimates. However, this assumption depends on the condition that training and test data come from similar distributions, which may not be true in real-world contexts. In fact, recent studies have already suggested that larger training datasets do not always yield better predictions~\cite{Leevy}. In some cases, smaller but well-curated samples can outperform large datasets when sampling is done thoughtfully~\cite{chang2009national}.

In the common situation where historical data is used to train a machine learning model to predict future outcomes, the ``more data is better'' assumption means a preference for including more data from earlier time periods in addition to recent data to generate predictions for a given future time point. 
While this strategy aims to improve model robustness, it may introduce outdated patterns that divert from more recent data points. This divergence is known as distribution shift in statistics and machine learning research~\cite{quinonero2009dataset}. Two types of distribution shift are commonly identified: \textit{covariate shift}, referring to changes in the distribution of input features, and \textit{concept shift}, referring to changes in the relationship between inputs and outcomes. Rigorous evaluation of the types and consequences of temporal distribution shifts is of practical importance, due partly to their potential negative impacts on model performance and partly to storage and computing costs associated with expanding training data.

Beyond achieving high model performance, predictions of machine learning applications in social contexts also ought to be algorithmically fair~\cite{mehrabi2021survey,kizilcec2021algorithmicfairnesseducation}, i.e., models perform equitably across different sociodemographic groups when making individual-level predictions. While fairness has been a common component of machine learning research and practice, intersectionality, which considers the compound and unique challenges faced by individuals with multiple marginalized identities, adds complexity to common fairness considerations~\cite{kong2022are}. Moreover, distribution shifts can be intertwined with fairness challenges, as different social groups may experience different types and rates of distribution shifts in their data.

Motivated by these concerns, this study systematically examines the impact of training data expansion under temporal distribution shifts on the performance and fairness of machine learning models by addressing the following research questions:
\begin{enumerate}
    \item How does expanding the historical training window affect model performance under varying degrees and types of temporal data distribution shifts?
    \item As the training window expands, how do varying degrees and types of temporal data distribution shifts explain any resulting performance degradation?
    \item As the training window expands, how do unequal temporal distribution shifts across groups affect model fairness?
\end{enumerate}

Our work is expected to contribute to prior research and practice in multiple ways. 
First, we advance the theoretical and empirical understanding of temporal stability of machine learning models through the lens of distribution shifts. Second, we provide empirical evidence for responsible AI by linking temporal distribution shifts to intersectional algorithmic fairness, revealing the complex ways in which data distributions shape fairness outcomes across intersectional groups.  
Third, we present a reproducible simulation and evaluation framework for expanding-window training under temporal data distribution shifts, with the potential to guide practitioners in data collection and model maintenance for machine learning applications.

\section{Related Work}
\subsection{Distribution Shift}
In supervised machine learning, data distribution shift refers to the phenomenon that the data distribution the model is trained on differs from the data distribution the model is tested on. Two commonly studied types of data distribution shift that influence model performance are covariate shift and concept shift, each reflecting a distinct mismatch between the training and testing distributions~\cite{kouw2019introductiondomainadaptationtransfer}:
\begin{itemize}
    \item \textbf{Covariate shift} refers to changes in the marginal distribution of the input features $P(X)$, while the conditional distribution $P(Y | X)$ remains invariant.
    \item \textbf{Concept shift} refers to changes in the conditional distribution $P(Y | X)$, indicating that the underlying relationship between features and labels changes.
\end{itemize}
A wide range of methods has been proposed for detecting data distribution shifts. For covariate shift, detection techniques involve statistical divergence measures (e.g., Kullback-Leibler divergence~\cite{csiszar1975i}, Maximum Mean Discrepancy~\cite{JMLR:v13:gretton12a}) or hypothesis testing procedures (e.g., Kolmogorov-Smirnov test~\cite{marsaglia2003evaluating}) to compare $P_{\text{train}}(X)$ and $P_{\text{test}}(X)$.
Detecting concept shift is more challenging because it involves estimating changes in $P(Y |X)$, which is not directly observable. Common approaches include model-based methods that track degradation in model performance metrics~\cite{svm} or using error-driven drift detectors such as the Drift Detection Method (DDM)~\cite{gama2004learning}. However, because our goal is to examine the relationship between data distribution shifts and model performance, using methods that infer shift from the evaluated model’s own errors or drift statistics introduces a circular dependency, i.e., the model defines the shift signal, and that signal is then used to explain the model’s performance. Therefore, we employ a nonparametric, model-agnostic approach based on \textit{k-nearest neighbors (kNN)} to estimate shifts in $P(Y | X)$ for the empirical study, which is introduced in detail in later sections.

Beyond detection, various strategies have been proposed to mitigate the adverse effects of distribution shifts, including instance reweighting, ensemble learning, and domain-invariant feature learning~\cite{azarkesht2022instance,bifet,lu2022domaininvariantfeatureexplorationdomain}. Recent work has also focused on diagnosing model performance degradation attributable to different types of distribution shifts~\cite{cai2023}. While the study provides a framework to examine performance drops between a single source–target pair, our temporal setting considers a sequence of source–target pairs generated by expanding the training window over time.

\subsection{Predictive Analytics in Education}
Predictive analytics has become a widely adopted approach in education. Institutions leverage large-scale data and advanced machine learning techniques to inform decision making~\cite{beaulac2019predicting}. One notable application is the development of early warning systems, which aim to identify students at risk of academic failure or dropout. These systems typically utilize behavioral, academic, and administrative data to generate timely risk predictions, which allow institutions to implement targeted interventions that support student success~\cite{berens2019early}.

Despite the growing adoption of predictive models, relatively little research has examined how the choice and scope of training data influence model performance, particularly under conditions of data distribution shift. Prior work has shown that model outcomes can be highly sensitive to analytical design decisions, such as variable selection, preprocessing, and modeling techniques~\cite{multiverse}. Beyond model specification, the alignment between training and testing data is also critical. A study on transfer learning in educational predictive modeling found that contextual information can be helpful in guiding model selection; in particular, more similar pairs of source and target institutions tend to yield better transfer model performance~\cite{lak25}. Other research has examined how the predictive value of features evolves over time and varies across student groups, finding that predictors that are informative at earlier stages can lose importance as new information becomes available in later periods~\cite{lak2024}. Another relevant study explored how data distribution shifts during the COVID-19 pandemic affected the performance of retention prediction models and found that, while imperfect, predictive models can still yield useful insights under certain conditions~\cite{covid}. Extending this line of work, we examine how the interaction between temporal distribution shifts and the expanding use of historical training data shapes the performance and fairness of predictive analytics.

\subsection{Algorithmic Fairness and Intersectionality}
Algorithmic fairness refers to the principle that machine learning models should yield equitable outcomes across diverse demographic groups~\cite{mehrabi2021survey}. A growing body of research has assessed whether models trained on the entire population produce systematically biased predictions for certain subgroups, particularly along dimensions such as race, gender, and socioeconomic status~\cite{kizilcec2021algorithmicfairnesseducation}. Another line of work has examined the role of protected attributes in model development, debating whether their inclusion can improve fairness without introducing additional harms~\cite{protected}. Research has also expanded from single-group fairness to intersectional fairness, which considers the compound disadvantages faced by individuals with intersectional marginalized identities~\cite{kong2022are}, an important perspective for uncovering disparities that single-axis analyses may overlook but one that remains underexplored in educational predictive modeling.

To address fairness concerns, some studies have proposed fairness-aware modeling approaches that incorporate fairness constraints during training~\cite{hu2020towards}. While current research has identified structural inequities ~\cite{barocas2023fairness} and data underrepresentation~\cite{bird2024algorithms} as sources of bias in educational prediction models, relatively little attention has been paid to the role of data distribution shifts in shaping algorithmic disparities. In this study, we address this gap by investigating how changes in data distributions contribute to algorithmic bias. Through this lens, we aim to provide novel explanations for observed disparities.

\section{Problem Setup and Methods}
\subsection{Prediction Task}
We focus on the common scenario of binary classification (e.g., at-risk or not) under expanding historical training windows. Specifically, predictive models are trained using data from prior time periods, with the training window growing larger as more historical data is added. This setup enables a systematic analysis of performance changes as the training window expands in reverse chronological order.
\subsection{Measuring Distribution Shift}
\subsubsection{Covariate Shift}
We quantify covariate shift separately for continuous and binary features.

For continuous features, we first standardize each variable and apply Principal Component Analysis (PCA) to reduce dimensionality. Let \( d_{\text{PCA}} \) denote the number of retained principal components. To measure distributional change along each component, we apply the Kolmogorov–Smirnov (KS) test between training and test sets. The KS statistic is defined as:

\[
D_j = \sup_z |F_{\text{train},j}(z) - F_{\text{test},j}(z)|
\]

where \( F_{\text{train},j} \) and \( F_{\text{test},j} \) are the empirical cumulative distribution functions (CDFs) of the \( j \)-th principal component in the training and test datasets. The value \( D_j \) captures the maximum difference between the two CDFs and reflects the marginal distribution shift. We then compute the average KS statistic across all components as a summary measure:

\[
\text{CovShift}_{\text{cont}} = \frac{1}{d_{\text{PCA}}} \sum_{j=1}^{d_{\text{PCA}}} D_j
\]

For binary features, we first calculate the absolute difference in positive class proportions:

\[
\Delta p_i = |p_{i,\text{train}} - p_{i,\text{test}}|
\]

where \( p_{i,\text{train}} \) and \( p_{i,\text{test}} \) denote the proportion of 1s in feature \( i \) in the training and test sets.

Next, we compute Cramér's V based on a \( 2 \times 2 \) contingency table:

\[
\text{Contingency Table}_i = 
\begin{bmatrix}
n_{\text{train},0}^{(i)} & n_{\text{train},1}^{(i)} \\
n_{\text{test},0}^{(i)} & n_{\text{test},1}^{(i)}
\end{bmatrix}
\]

where \( n_{g,v}^{(i)} \) is the count of group \( g \in \{\text{train}, \text{test}\} \) with value \( v \in \{0,1\} \). Cramér's V is calculated as:

\[
V_i = \sqrt{\frac{\chi^2 / n_{\text{total}}}{\min(r - 1, c - 1)}}
\]

where \( \chi^2 \) is the Chi-squared statistic from the table, \( n_{\text{total}} \) is the total number of observations, and \( r = c = 2 \) are the table dimensions.

The binary covariate shift is defined as the average of the delta proportions and Cramér’s V scores across all binary features:

\[
\text{CovShift}_{\text{bin}} = \frac{1}{2}\left( \frac{1}{d_{\text{bin}}} \sum_{i=1}^{d_{\text{bin}}} \Delta p_i + \frac{1}{d_{\text{bin}}} \sum_{i=1}^{d_{\text{bin}}} V_i \right)
\]
To synthesize continuous and binary covariate shifts into one metric, we calculate a unified covariate shift score as their arithmetic mean:

\[
\text{CovShift}_{\text{unified}} = \frac{\text{CovShift}_{\text{cont}} + \text{CovShift}_{\text{bin}}}{2}
\]

\subsubsection{Concept Shift}
In the simulation setting, we have access to the ground-truth coefficients $\{\beta\}$ that generate the data. We therefore measure concept shift in an oracle, model-agnostic manner by comparing the conditional label distributions induced by the training and testing coefficients. Specifically, let $\beta_{\text{train}}$ and $\beta_{\text{test}}$ denote the training and testing coefficient vectors, and let $Q_{\beta}(\cdot\mid x)$ be the conditional label distribution of $Y$ given $X=x$ under coefficients $\beta$. For test covariates $\{x_i\}_{i=1}^n$, define
\[
\operatorname{ConceptShift}
=\frac{1}{n}\sum_{i=1}^{n}
\operatorname{JS}\!\big(
Q_{\beta_{\text{train}}}(\cdot\mid x_i)\,,\,
Q_{\beta_{\text{test}}}(\cdot\mid x_i)
\big),
\]
where $\operatorname{JS}$ denotes the Jensen–Shannon divergence, a symmetric and bounded measure of dissimilarity between probability distributions. This yields a single scalar summarizing how much the conditional label distributions implied by $\beta_{\text{train}}$ and $\beta_{\text{test}}$ differ. By design, this pointwise comparison does not depend on the marginal feature distribution $P(X)$; hence, the score is unaffected by covariate shift.

However, in real-world data, the true $\{\beta\}$ are unobserved, and using the oracle metric above is not feasible in practice. Therefore, in our empirical study, we adopt a model-agnostic and non-parametric approach based on \( k \)-nearest neighbors (kNN) to estimate the conditional distribution. This method allows us to capture local changes in the relationship between features and the target variable without imposing strong functional assumptions.

The rationale for using kNN lies in its locality: by averaging outcomes over nearby points in the feature space, the method approximates the conditional expectation \( \mathbb{E}[Y | X = x] \). Unlike parametric models that assume a fixed functional form, kNN can flexibly adapt to complex and potentially non-stationary relationships between features and outcomes. 

Specifically, we first project the standardized input features into a lower-dimensional space using Principal Component Analysis (PCA) to improve the stability and efficiency of distance-based computations. In this reduced feature space, we estimate the conditional probability \( \hat{p}(x) = P(Y = 1 | X = x) \) for each observation by averaging the observed labels of its \( k \)-nearest neighbors.

Let \( \hat{p}_{\text{train}} \) and \( \hat{p}_{\text{test}} \) denote the estimated conditional probabilities from the training and test sets, respectively. We then compute the concept shift score as:

\[
\text{ConceptShift}_{\text{JS}} = \text{JS}(\hat{p}_{\text{train}}, \hat{p}_{\text{test}})
\]

While our kNN-based approach provides a flexible, non-parametric estimate of the conditional distribution \( P(Y|X) \), it is inherently sensitive to changes in the input distribution \( P(X) \). That is, when covariate shift is present, the neighborhoods identified by the kNN algorithm may differ between training and test sets, even if the underlying conditional relationship remains stable. As a result, estimated differences in \( P(Y|X) \) may conflate genuine concept shift with distortions introduced by covariate shift.

To address this issue, we perform a residualization procedure to isolate the portion of concept shift that cannot be explained by covariate shift alone. Specifically, we regress the raw concept shift scores on the unified covariate shift score:

\[
\text{ConceptShift}_{\text{JS}} = f(\text{CovShift}_{\text{unified}}) + \varepsilon
\]

where \( f(\cdot) \) is the random forest function that captures both linear and non-linear dependencies. The residual term \( \varepsilon \) represents the unexplained component.
The residualized concept shift metric is defined as
\[
\text{ConceptShift}_{\text{JS, resid}} = \text{ConceptShift}_{\text{JS}} - \widehat{f}(\text{CovShift}_{\text{unified}})
\]

Higher values indicate larger discrepancies in the conditional distributions.

\subsection{Performance Evaluation}
In binary predictions, various performance metrics have been proposed in the literature. Common metrics derived from the confusion matrix (e.g., accuracy, precision, recall) require the selection of a fixed decision threshold. However, our setting involves expanding training windows and temporal data distribution shifts, which makes it challenging to determine a stable or meaningful threshold across time windows. Therefore, we adopt the \textbf{Area Under the Receiver Operating Characteristic Curve (AUC)} as our primary evaluation metric. AUC is threshold-independent and summarizes model performance across all possible classification thresholds.

Formally, AUC is defined as:
\[
\text{AUC}(f(\theta)) = \int_{0}^{1} \text{TPR}(\text{FPR}^{-1}(t)) \, dt
\]
where \( t \) denotes a decision threshold, \( f_t(\theta, x) = 1 \) if \( f(\theta, x) \geq t \), and TPR and FPR represent the true positive rate and false positive rate, respectively. AUC values range from 0 to 1, with higher values indicating better discriminative ability; an AUC of 0.5 corresponds to random guessing. 

To assess fairness across demographic subgroups and capture intersectional disparities, we use the \textbf{AUC Gap}~\cite{Gardner_2023} as a group fairness metric. The AUC Gap is defined as:
\[
\max_{g, g' \in \mathcal{G}} \left| \mathbb{E}_{\mathcal{D}_k} \left[ \text{AUC}(f_{\theta} \mid \mathcal{D}_{k,g}) \right] - \mathbb{E}_{\mathcal{D}_k} \left[ \text{AUC}(f_{\theta} \mid \mathcal{D}_{k,g'}) \right] \right|
\]
where \( \mathcal{D}_{k,g} \) and \( \mathcal{D}_{k,g'} \) denote the evaluation data restricted to subgroups \( g \) and \( g' \) respectively. This metric captures the worst-case difference in AUC across subgroups and serves as a conservative indicator of fairness degradation under distribution shifts.

\subsection{Statistical Analysis}

To examine the relationship between data distribution shift and model outcomes, we employ regression analysis as our primary analytical method. Specifically, we use linear regression models to examine how covariate shift and concept shift relate to model performance and fairness.

For model performance:
\begin{equation}
\text{AUC}_i = \beta_0 + \beta_1 \cdot \delta_i + \beta_2 \cdot \theta_i + \beta_3 \cdot (\delta_i \times \theta_i) + \mathbb{I}_{\text{emp}} \cdot \gamma_{j[i]} + \epsilon_i,
\label{eq:performance}
\end{equation}
where \(\text{AUC}_i\) denotes model performance for unit \(i\). The variables \(\delta_i\) and \(\theta_i\) represent covariate shift metric and concept shift metric, respectively. \(\mathbb{I}_{\text{emp}}\) is a binary indicator that equals 1 for empirical study and 0 for simulation study, activating school fixed effects \(\gamma_{j[i]}\) in the empirical study to account for institutional heterogeneity.

To assess model fairness, we examine the relationship between shift disparities across demographic groups and the AUC gap
\begin{equation}
\text{AUCGap}_i = \beta_0 + \beta_1 \cdot \Delta_i + \beta_2 \cdot \Theta_i + \mathbb{I}_{\text{emp}} \cdot \gamma_{j[i]} + \epsilon_i,
\label{eq:fairness}
\end{equation}
where \(\text{AUCGap}_i\) measures fairness disparity for unit \(i\). The terms \(\Delta_i\) and \(\Theta_i\) represent the (max-min) gaps in covariate and concept shift across demographic groups.

\section{Simulation Study}
\subsection{Simulation Process}

We design a simulation framework to examine how temporal covariate shift and concept shift affect model performance and fairness under an expanding training window setting. Below, we describe how we simulate each type of shift and outline the experimental scenarios used in our study.

\subsubsection{Covariate Shift}

To simulate temporal covariate shift, we allow the marginal distribution \( P(X) \) to vary across time while keeping \( P(Y | X) \) fixed. Specifically, we apply a time-dependent mean shift to the continuous features. For each continuous feature \( x_j \), its mean at time \( t \) is defined as:
\[
\mu_j^{(t)} = \mu_j^{(0)} + \delta_j \cdot \alpha_t^X
\]
where \( \delta_j \) is a feature-specific drift direction, \(\alpha_t^X \in [0,1]\) is a progression parameter that controls the magnitude of the shift over time, and \(\mu_j^{(0)}\) is the baseline mean.  

Binary features also shift in marginal proportions by adjusting their Bernoulli probabilities over time. For each binary feature \( x_j \), the probability of success at time \( t \) is:
\[
\pi_j^{(t)} = \pi_j^{(0)} + \rho_j \cdot \alpha_t^X
\]
where \( \pi_j^{(0)}\) is the baseline probability, \(\rho_j\) determines the rate of change, and probabilities are clipped to remain strictly between 0 and 1.

\subsubsection{Concept Shift}

To simulate temporal concept shift, we allow the conditional relationship \( P(Y | X) \) to evolve over time. Specifically, we define two sets of coefficients \( \beta^{(0)} \) and \( \beta^{(1)} \), representing the start and end states of the underlying data-generating process. For each year \( t \in \{1, \dots, T\} \), we interpolate linearly between them:

\[
\beta^{(t)} = (1 - \alpha_t)\beta^{(0)} + \alpha_t\beta^{(1)}, \quad \text{with} \quad \alpha_t = \frac{t-1}{T-1}
\]

Given features \( X_t \), the log-odds of the binary outcome are computed as \[
\text{logit}(P(Y = 1 | X)) = X_t \cdot \beta^{(t)}
\]

\subsubsection{Simulation Methods}

To implement the simulation scenarios, we generate synthetic tabular datasets with a consistent feature structure across all time periods. 

Continuous features are independently sampled from Gaussian distributions with fixed or time-varying means, depending on the covariate shift condition. Similarly, binary features are drawn from independent Bernoulli distributions with fixed or drifting probabilities, depending on whether binary covariate shift is introduced.

For each time period \( t \in \{1, \dots, 50\} \), we independently generate a dataset \( \mathcal{D}_t = \{(X_t^{(i)}, Y_t^{(i)})\}_{i=1}^{n} \) consisting of 5,000 samples, where each feature vector \( X_t^{(i)} \in \mathbb{R}^{19} \) concatenates 15 continuous and 4 binary features. The corresponding binary label \( Y_t^{(i)} \in \{0,1\} \) is sampled from a Bernoulli distribution, with the success probability determined by the logistic model at time \( t \).

\subsection{Simulation Scenarios}
\subsubsection{Model Performance}

To address RQ1 and RQ2, we simulate four scenarios to examine how covariate and concept shifts influence model performance under expanding training windows. In each scenario, data is generated from a common process across the entire population, without group-specific variation. Predictive performance is assessed using AUC on a fixed test time period (i.e., the most recent time period), while the training set is progressively expanded by incorporating additional data from earlier time periods.

\begin{itemize}
    \item \textbf{Scenario (A): No Shift} — Both the feature distribution \( P(X) \) and the conditional relationship \( P(Y|X) \) remain stationary over time. This serves as a baseline to observe performance under temporal stability.

    \item \textbf{Scenario (B): Covariate Shift Only} — The marginal distribution \( P(X) \) changes gradually over time, while the conditional distribution \( P(Y | X) \) remains fixed. 

    \item \textbf{Scenario (C): Concept Shift Only} — The conditional relationship \( P(Y | X) \) changes over time through smooth transitions in model coefficients, while the feature distribution remains constant. 

    \item \textbf{Scenario (D): Combined Shift} — Both \( P(X) \) and \( P(Y | X) \) change over time. This setting reflects the compounded effects of simultaneous covariate and concept shifts.
\end{itemize}

In all cases, the logistic regression model is trained using data from an expanding window ending at time \( t-1 \), and evaluated on a fixed test year \( t \).

\subsubsection{Model Fairness}

To address RQ3, we extend the simulation framework to incorporate group-specific distribution shifts. Our goal is to assess whether covariate or concept shifts lead to disproportionate performance degradation for certain demographic groups, with particular attention to intersectional subgroups.

We consider two fairness-specific simulation scenarios:

\begin{itemize}
    \item \textbf{Scenario (E): Single-Group Shift} — One binary group (e.g., group = 1 for a demographic attribute) experiences either covariate shift or concept shift over time, while the other group remains stationary. This setting allows us to measure whether group-based shifts lead to growing AUC gaps between groups.

    \item \textbf{Scenario (F): Double-Group Shift} — Two binary demographic attributes, G1 and G2, jointly define four intersectional subgroups. One subgroup from each attribute independently undergoes concept shift. We manipulate the \textit{direction} of these shifts—either aligned (same direction) or opposed. This setup enables us to evaluate how the alignment of shifts across demographic dimensions affects fairness, and whether intersectional subgroups suffer more pronounced fairness degradation when exposed to conflicting versus reinforcing shift patterns.
\end{itemize} 

In both cases, the logistic regression model is trained using full data from an expanding window ending at time \( t-1 \), with performance assessed separately for each subgroup.

\subsection{Results}
\subsubsection{Model Performance}
\begin{figure*}[h]
    \centering
    \includegraphics[width=0.9\linewidth]{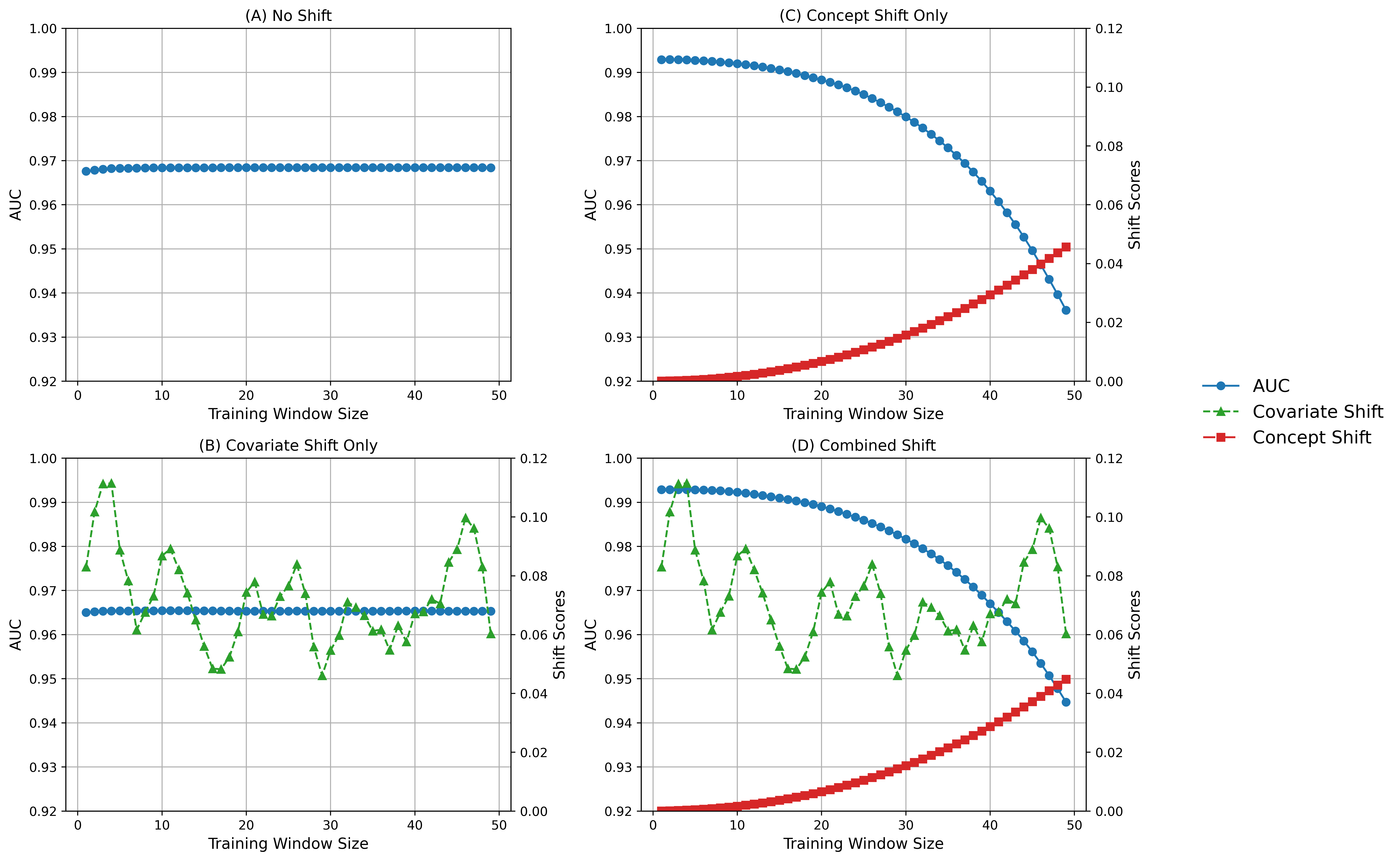}
    \caption{Model performance and data distribution shift metrics across training window sizes. Each panel shows how AUC (left axis) and shift scores (right axis) change as the training window expands in four simulated scenarios: (A) No shift, (B) Covariate shift only, (C) Concept shift only, and (D) Combined shift. Scales are not directly comparable across metrics; only within-metric trends matter.}
    \label{fig:shift_analysis_simulation}
\end{figure*}

Figure~\ref{fig:shift_analysis_simulation} presents model performance and data distribution shift metrics across the four simulated scenarios introduced above. To address RQ1, we examine how expanding the training window size affects model performance under different types of temporal shift.

In the \textit{No Shift} scenario, model performance remains largely flat as the training set increases, with a small improvement at the beginning. In the \textit{Covariate Shift Only} scenario, although the covariate shift metric varies over time, model performance remains stable. Under the \textit{Concept Shift Only} scenario, model performance is stable at first but then consistently declines as the training window expands. Lastly, in the \textit{Combined Shift} scenario, we also observe performance degradation over time, but the magnitude is less severe compared to the \textit{Concept Shift Only} scenario.

To address RQ2, we examine how different types of data distribution shifts relate to changes in model performance as described in Equation~\ref{eq:performance}. The full regression results are presented in the Appendix\footnote{Appendix is available at https://github.com/AEQUITAS-Lab/Distribution-Shift-AIES-2025}. Across all simulated scenarios, we find that concept shift has a consistently negative and statistically significant association with model performance degradation. In \textit{Concept Shift Only} scenario, the coefficient for concept shift is $\beta = -0.4040$ ($p < .001$). In the \textit{Combined Shift} scenario, the coefficient for concept shift is $\beta = -0.6651$ ($p < .001$). Covariate shift alone, as examined in \textit{Covariate Shift Only} scenario, shows no significant relationship with model performance ($\beta = -0.0001$, $p = 0.502$), consistent with the flat AUC trend observed under that condition. In addition, in the \textit{Combined Shift} scenario, covariate shift still shows no significant relationship with model performance ($\beta = -0.0205$, $p = 0.154$). The interaction between concept shift and covariate shift is not statistically significant, suggesting that their joint occurrence does not affect model performance beyond the impact of each shift individually; thus, concept shift is the primary driver of performance degradation in our simulations. One limitation of our simulation design is that covariate and concept shifts are generated independently; in real-world settings, these shifts may interact in more complex and intertwined ways.

\subsubsection{Model Fairness}
\begin{figure*}[h]
    \centering
    \includegraphics[width=0.9\linewidth]{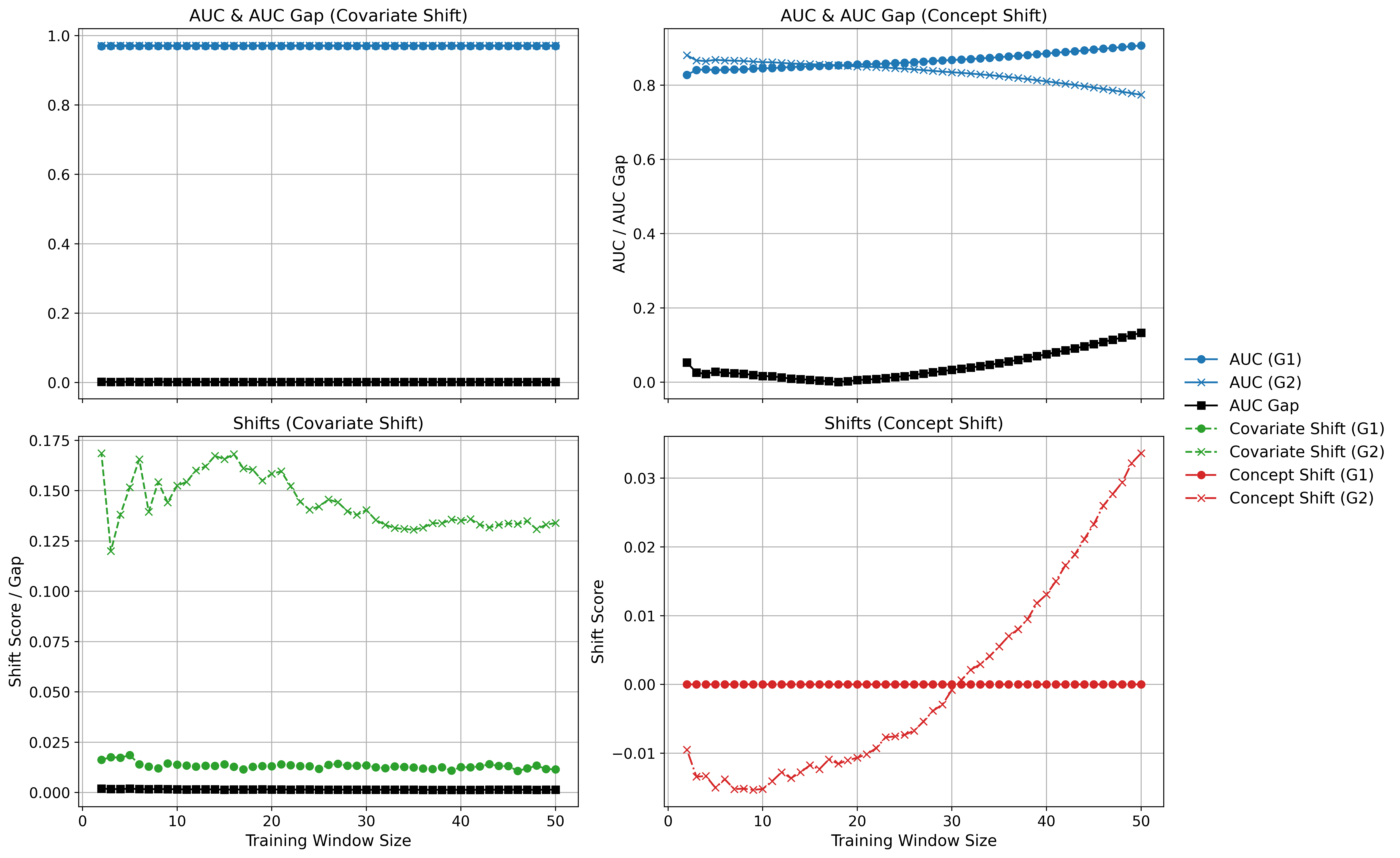}
    \caption{Subgroup model performance and shift metrics across training window sizes. Each panel presents AUC and shift measurements for two demographic subgroups under simulated distribution shift conditions. The left column corresponds to covariate shift only, and the right column to concept shift only. The top row shows subgroup AUCs and AUC gap, while the bottom row displays group-specific shift metrics. Scales are not directly comparable across metrics; only within-metric trends matter.}
    \label{fig:auc_auc_gap_cov}
\end{figure*}
To address RQ3, Figure~\ref{fig:auc_auc_gap_cov} presents the results for the \textit{Single-Group Shift} scenario. In the Concept Shift Only condition, we observe that the subgroup exposed to the shift exhibits a clear performance decline, while the unshifted group shows a gradual performance improvement. This increase occurs because the global model’s coefficients are pulled closer to the stable subgroup’s true decision boundary as the training window expands. As a result, the fairness metric—\textit{AUC Gap} between the two groups—increases as the training window expands. 

In contrast, in the Covariate Shift Only condition, even though one group undergoes greater covariate distributional shift, model performance remains largely unaffected and the \textit{AUC Gap} stays flat. This finding indicates that covariate shift alone does not necessarily compromise predictive parity.

To formally quantify the effects, we conducted regression analyses based on Equation~\ref{eq:fairness}. The group-level difference in concept shift metric was a significant predictor of fairness degradation, with a coefficient of $\beta = 2.5089$ ($p < .001$). This result implies that larger inter-group disparities in concept shift are associated with greater disparities in model performance.

\begin{figure*}[h]
    \centering
    \includegraphics[width=0.9\linewidth]{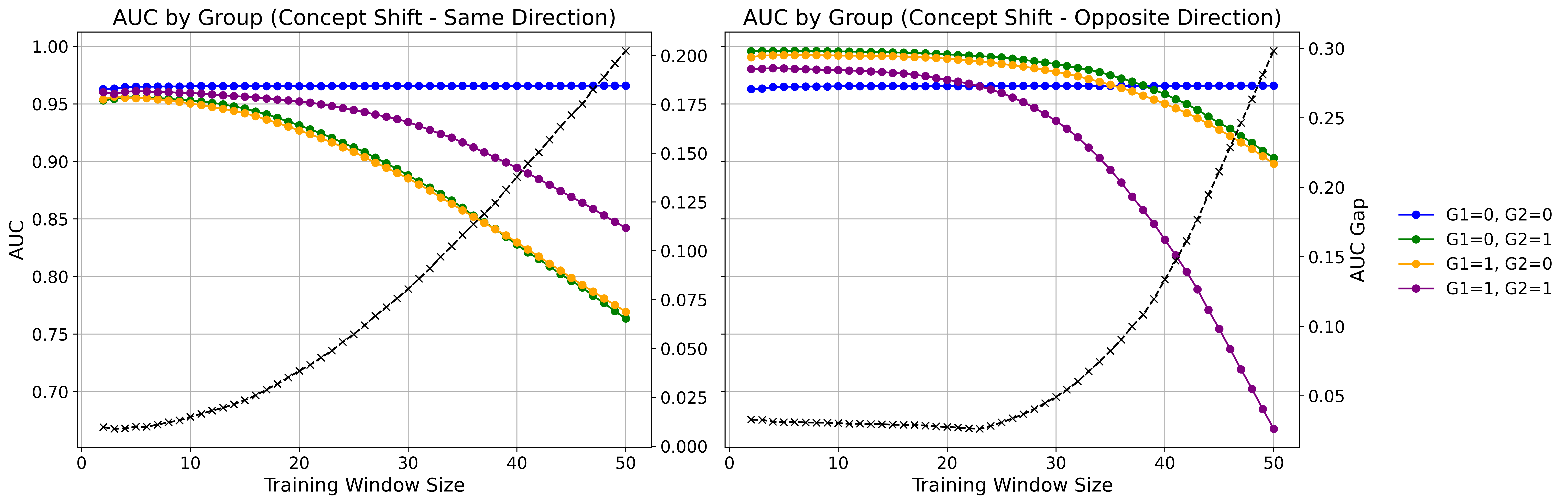}
    \caption{Subgroup AUC and fairness gap under concept shift in same and opposite directions. Each panel shows subgroup-specific AUCs (left axis) and the resulting AUC gap (right axis) across training window sizes. The left panel simulates concept shift in the same direction for all intersectional subgroups, while the right panel simulates concept shift in opposite directions for different subgroups. Lines represent intersectional groups defined by G1 and G2.}
    \label{fig:auc_inter}
\end{figure*}

We next examined the \textit{Double-Group Shift} scenario to assess how intersectional dynamics influence fairness outcomes. Figure~\ref{fig:auc_inter} illustrates the impact of concept shift on model fairness when two demographic dimensions are jointly shifted. The effects of intersectional shift are not merely the sum of individual group shifts. When the concept shift directions are aligned across groups, the intersectional group may experience attenuated disparity. In contrast, when shift directions are opposite, the intersectional disparities can become amplified, resulting in a larger AUC gap. 

\section{Empirical Case Study}
\subsection{Study Context and Data}
Our empirical study focused on predicting student retention in postsecondary education. The central technical objective is to predict first-year retention using the expanding training window approach. First-year retention is defined as whether a student who enters an institution for the first time in a Fall term subsequently re-enrolls at the same institution in the following Fall. This definition aligns with the standard used by the federal government~\cite{gardner2022persistence}.

Through an established research partnership, we obtained access to detailed administrative records from 23 community colleges located within a southern state in the United States. These records included a wide range of student-level data, such as demographic background and academic performance. Community colleges are two-year public institutions that play a critical role in the U.S. higher education system. They serve as a primary access point to postsecondary education for many students from underrepresented and underserved backgrounds, including low-income, first-generation, and minority students. Compared to four-year research universities, community colleges typically have lower retention and completion rates~\cite{USDOE_CommunityCollegeFacts}.

For this study, we restrict the sample to \textit{first-time, first-year} students who entered college during Fall terms from 2010 to 2021, resulting in a dataset of 1,307,789 students. Among these students, the overall composition comprises 56.7\% female students and 31.1\% underrepresented minority (URM\footnote{URM refers to students who identify as Black/African American, Hispanic/Latino, or American Indian, in line with institutional reporting practices in the United States.}) students. The largest intersectional subgroup is \textit{Non-URM Female} (37.5\%), followed by \textit{Non-URM Male} (31.4\%), \textit{URM Female} (18.1\%), and \textit{URM Male} (12.9\%). To ensure comparability across institutions, we constructed a shared data schema based on commonly available variables. For each college, we designate the most recent year (2021) as the fixed test set and construct expanding training windows by progressively incorporating data from earlier years. The shared data schema is documented in the Appendix.

\subsection{Results}
\subsubsection{Predictive Performance}
\begin{figure}[h]
    \centering
    \includegraphics[width=0.9\linewidth]{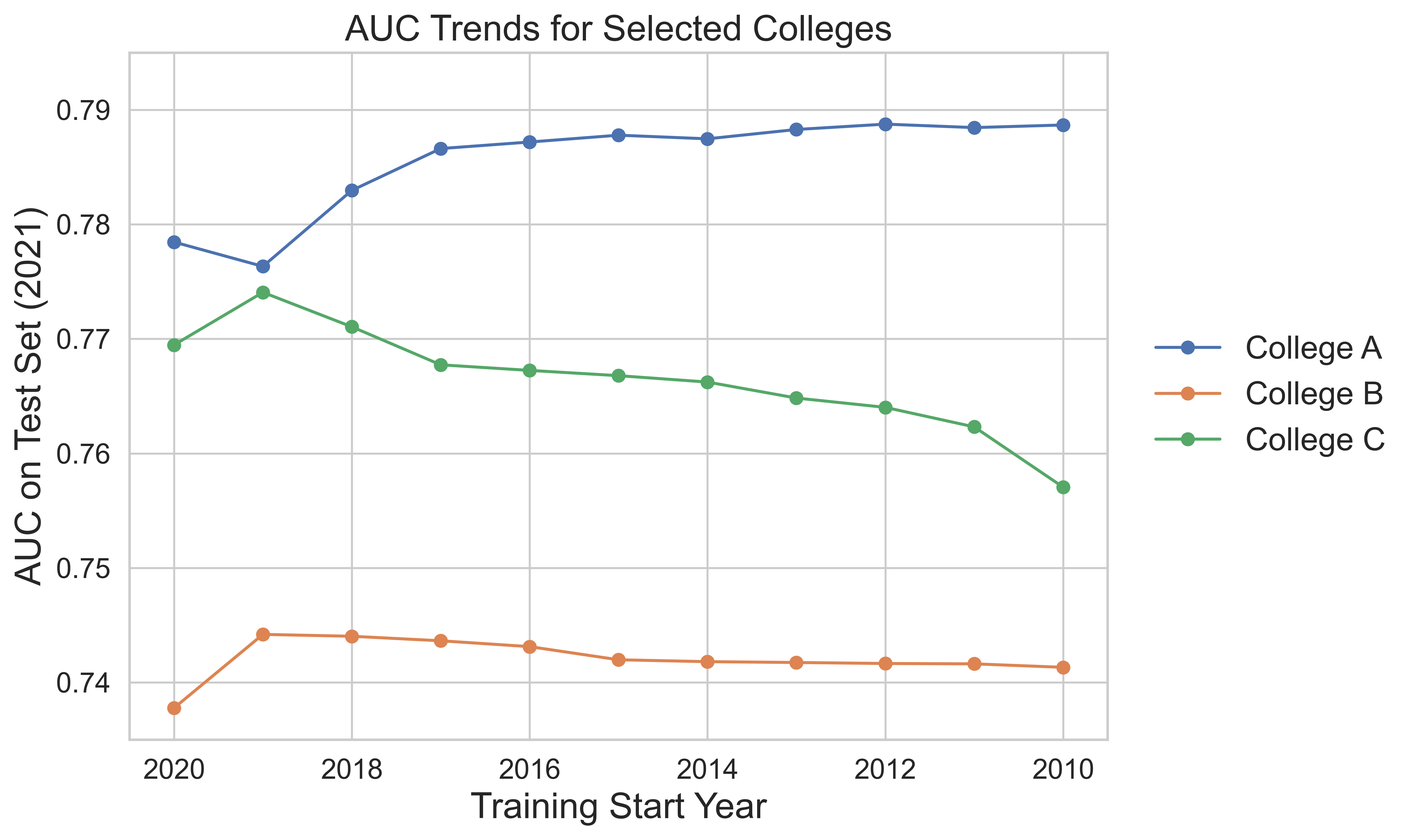}
    \caption{AUC trends by training start year for selected colleges. Model performance (AUC) on the 2021 test set is shown for three representative colleges, plotted against the training start year. Each curve corresponds to an expanding training window that begins in the indicated year and ends in 2020. Full results for all colleges are provided in the Appendix.}
    \label{fig:auc_time}
\end{figure}
Figure~\ref{fig:auc_time} illustrates representative performance trajectories for three colleges over time. Each line reflects AUC performance on the fixed test set (2021) as the training window expands. The selected colleges exemplify three common patterns observed across the full dataset. College B shows a relatively stable trend, suggesting that accumulating historical data has limited influence on retention prediction. College A demonstrates a steady but modest improvement, indicating that additional training history enhances model performance. College C follows a rise-then-decline pattern, where performance initially improves but eventually deteriorates as older data is added.
All remaining colleges conform to one of these three general patterns, with full results reported in the Appendix. These findings show that increasing training data is not universally beneficial. While some colleges benefit from an expanding training window, others see minimal gains or even performance degradation.

\begin{figure*}[h]
    \centering \includegraphics[width=0.9\linewidth]{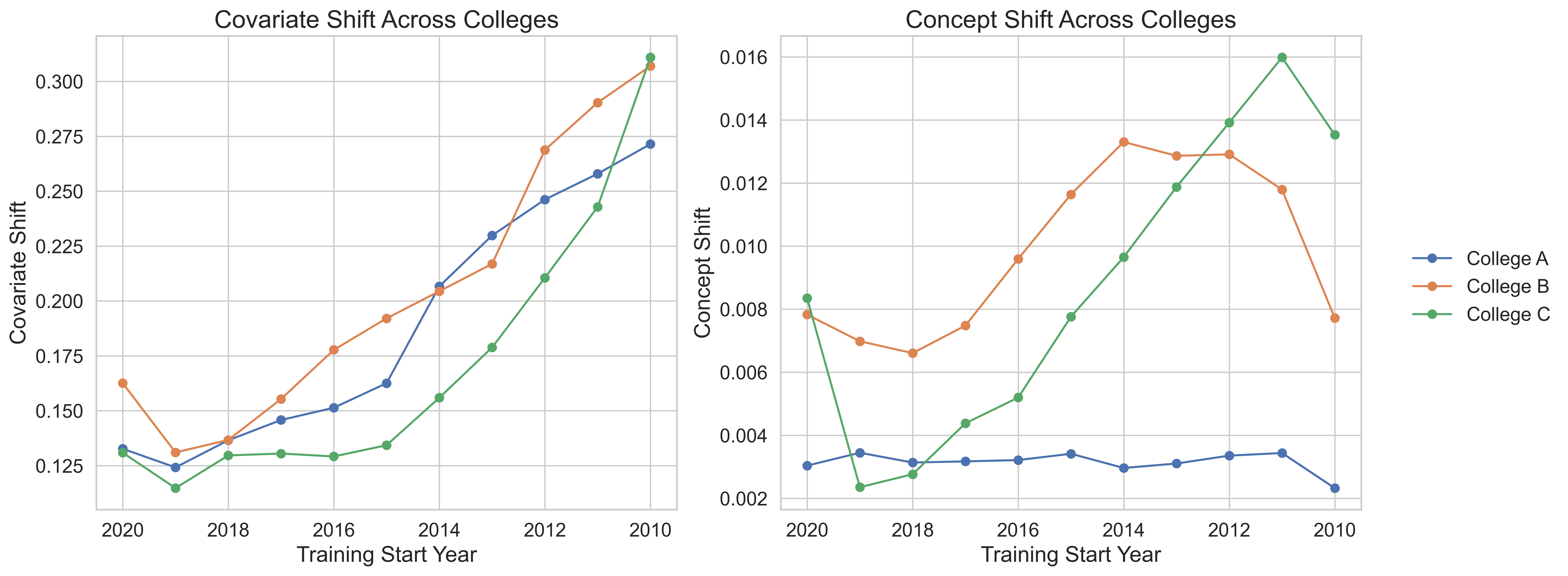}
    \caption{Covariate shift and concept shift by training start year for selected colleges. Magnitudes are computed between each expanding training window and the fixed 2021 test year for three representative colleges. Full results for all colleges are reported in the Appendix.}
    \label{fig:shift_trends}
\end{figure*}
Figure~\ref{fig:shift_trends} presents the covariate shift and concept shift trajectories for the same three exemplar colleges. Covariate shift trajectories are relatively consistent across colleges and show a steady increase as the training window expands. This pattern reflects gradual changes in the marginal distribution of input features over time. By contrast, concept shift trajectories exhibit more variation across colleges. Although the specific patterns differ, the overall trend is upward and indicates increasing divergence in the conditional relationship between features and outcomes as training windows expand. Full results for all colleges are available in the Appendix.

To examine how temporal shift influences predictive performance, we conduct a regression analysis on the pooled dataset across all colleges, following Equation~\ref{eq:performance}. At the full-sample level, none of the examined metrics show a statistically significant main effect on performance. One possible explanation is that the size of the training set conditions the observable impact of concept shift: larger training windows may amplify the effect of even modest shifts, whereas smaller windows may introduce high variance that obscures the influence of substantial shifts. Motivated by this consideration, we explore whether training size modulates the impact of concept shift on model performance. Specifically, we estimate the regression coefficient capturing the relationship between concept shift and performance across training sets of varying sizes and find that models trained on larger datasets exhibited a more stable and interpretable association between concept shift and predictive performance; detailed results are provided in the Appendix. To investigate this further, we partition the training sets into tertiles by size (Low, Medium, High) and re-estimated Equation~\ref{eq:performance} separately for each group. The results reveal that the coefficient for concept shift became increasingly negative and statistically significant as the training set expanded. In the largest tertile, concept shift had a statistically significant and pronounced negative effect on performance ($\beta = -0.5412$, $p = .003$). These findings suggest that the adverse effects of concept shift become more detectable when models are trained on sufficiently large historical datasets. The effect of covariate shift was more mixed: it was statistically significant in both the smallest and largest groups, but the corresponding coefficients were relatively small, indicating a modest and inconsistent influence on performance. Notably, the interaction between concept and covariate shift became increasingly positive and statistically significant with larger training sets. These results indicate that the co-occurrence of covariate and concept shift can lead to compounded performance degradation.

\subsubsection{Algorithmic Fairness}

\begin{figure*}[h]
    \centering
    \includegraphics[width=0.9\linewidth]{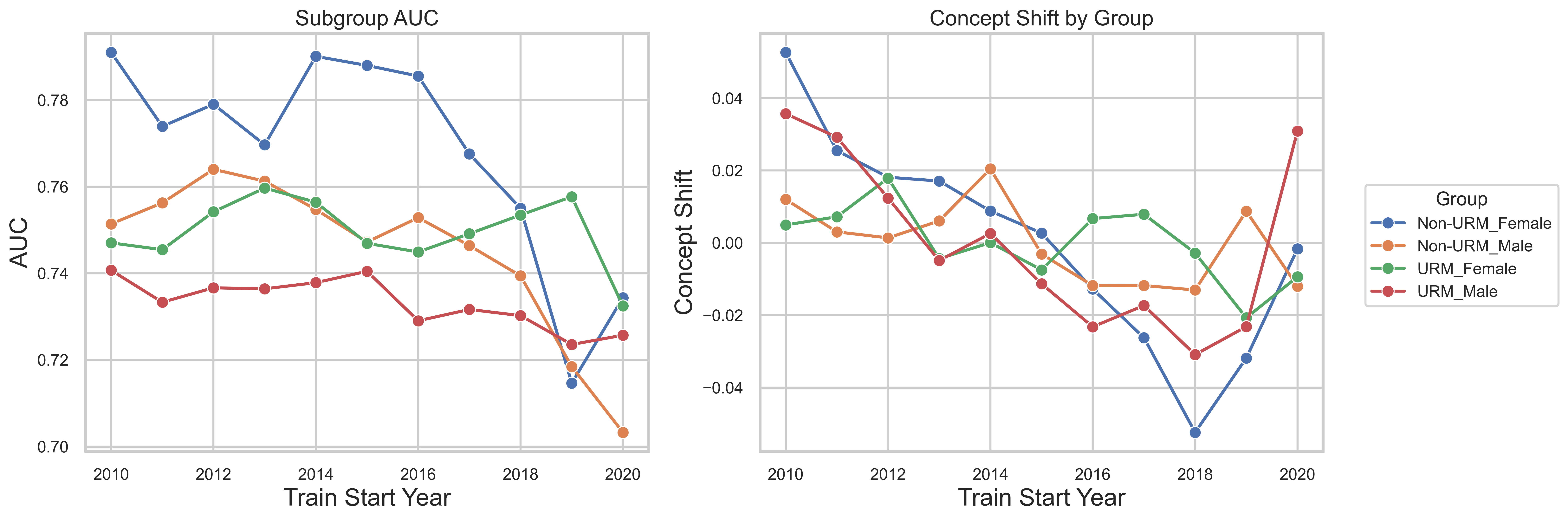}
    \caption{Subgroup AUC and concept shift by training start year for one college. The left panel shows AUC scores for four intersectional subgroups defined by URM status and gender across different training start years. The right panel shows the corresponding subgroup-specific concept shift metrics relative to the fixed 2021 test year. Full results for all colleges are provided in the Appendix.}
    \label{fig:intersectional_shift_perf}
\end{figure*}

We examine model fairness by constructing four intersectional subgroups based on gender and underrepresented racial minority (URM) status. The model is trained on the full dataset, and subgroup-specific performance and shift metrics were computed using a fixed test year. Figure~\ref{fig:intersectional_shift_perf} illustrates representative patterns of both distribution shift and AUC across the four groups. We do not observe a consistent pattern in which any particular group experienced uniformly greater exposure to shift or worse model performance across colleges. This is consistent with our simulation findings that the fairness impact on intersectional groups cannot be understood as a simple additive combination of single-group effects.

To more formally assess the relationship between distribution shift and model fairness, we apply the regression framework outlined in Equation~\ref{eq:fairness}. Our analysis shows that the gap in concept shift across groups is a statistically significant predictor of the AUC gap ($\beta = 0.5558$, $p = .002$), suggesting that fairness disparities are more likely to emerge when certain groups undergo substantially different degrees of concept shift. In contrast, the gap in covariate shift is not significantly associated with the AUC gap ($\beta = 0.0055$, $p = .916$). These findings are consistent with our simulation results, which indicate that larger inter-group differences in concept shift lead to more pronounced disparities in model performance.

\section{Discussion and Conclusion}
In this study, we examine the issue of predictive analytics under expanding temporal training windows, where earlier historical data is progressively incorporated to train models to predict future outcomes. Using a simulation study and a large-scale empirical analysis in the education sector, we present how the expansion of historical training data interacts with temporal data distribution shifts, specifically covariate and concept shift, to impact both model performance and fairness.

Both simulation and empirical results challenge the conventional assumption that simply increasing the volume of training data from the past would improve model performance. In dynamic environments characterized by distributional changes in historical data, this strategy can result in diminishing returns or even performance degradation. These findings underscore that predictive effectiveness depends not only on the quantity of data, but also on its temporal relevance and alignment with evolving patterns.

Beyond performance, we observe that uneven exposure to temporal concept shifts across sociodemographic groups leads to disparities in subgroup predictive performance. Importantly, among intersectional groups, the effects of multiple identities interact in ways that cannot be understood as a simple sum of individual group effects. These observations align with prior studies highlighting that models that are fair with respect to individual attributes like race or gender may still exhibit unfairness at their intersections~\cite{inter}. As a consequence, fairness-aware modeling should move beyond static group comparisons to account for both temporal and intersectional variation in data conditions.

Our study advances understanding of how distribution shifts shape the performance and fairness of predictive models under expanding training windows. By disentangling the distinct contributions of covariate and concept shift, we demonstrate that the benefits of adding more historical data depend on the nature and magnitude of underlying shifts. We also introduce a reproducible simulation framework capable of generating controlled and decoupled shift scenarios. Furthermore, we extend fairness analysis by showing that inter-group disparities in concept shift can be a key driver of fairness degradation in predictive modeling.

This study has several practical implications. For model developers, our findings emphasize the importance of evaluating distributional changes in training data when expanding historical datasets. While we do not have the capacity to directly identify the point at which additional data no longer enhances model performance, our results demonstrate that temporal distribution shifts can affect performance, which underscores the need for reasonable training dataset selection to reduce potential degradation and to avoid unnecessary data accumulation, storage needs, and computational costs. For institutional researchers, our findings highlight the importance of monitoring subgroup-level performance longitudinally and examining whether emerging disparities are associated with uneven exposure to shifts in the underlying population or behavioral patterns. For fairness researchers, our work extends beyond auditing model outcomes to understanding why models become less fair by offering a data distribution shift perspective as an explanatory lens.

Our study also has several limitations. First, while the simulation design offers strong control over the shift processes, real-world data may involve more complex and interacting shifts, as well as unobserved patterns, that influence both model performance and fairness, as seen in our empirical analysis. Second, although our kNN-based method provides a non-parametric, model-agnostic way to estimate concept shift, it is not without limitations and may be influenced by factors such as feature scaling, high dimensionality, or sparsity in certain regions of the feature space. Future work could explore purely statistical approaches to measuring concept shift that further reduce dependence on specific modeling. Third, although we demonstrate that temporal distribution shift can affect model performance, we do not yet provide an accurate method for identifying the exact point at which additional historical data ceases to be useful. Future research should seek to quantify this threshold and investigate how concept shift interacts with training data scope in ways that make certain historical data detrimental rather than beneficial. Finally, although our dataset spans a large number of institutions, the analysis remains contextually bounded, as it draws from a single state system and focuses on a specific predictive task. Extending this work to other domains that rely on historical data to predict future outcomes—such as employment forecasting, financial risk modeling, or public health monitoring—would help assess the generalizability of our findings.

\section{Acknowledgements}
This work was supported by funding from the Learning Engineering Virtual Institute through the Fairness Analysis and Transfer Learning Hub. We extend our gratitude to Dr. Catherine Finnegan for her invaluable data support.

\bibliography{aaai25}
\end{document}